\documentclass[aps,prb,floatfix,reprint]{revtex4-2}
\usepackage[utf8]{inputenc}
\usepackage{amsmath,amsbsy,amssymb,amsfonts}
\usepackage{graphicx, color}
\usepackage{esint}
\usepackage{empheq}
\usepackage{hyperref}
\usepackage{appendix} 

\makeatletter

\usepackage{subfigure}
\usepackage{soul}
\usepackage{xcolor}
\usepackage{soul}
\usepackage{amsthm}
\usepackage{color}
\usepackage{bm}
\usepackage{mathrsfs}
\usepackage{graphics}
\usepackage{lipsum}

\newcommand{\bb}[1]{\mathbb{#1}}

\makeatother

\begin{document}

\title{In-situ adjoint protocols for nonlinear PT-symmetric self-optimizing machines}

\author{Zheming Li$^{1}$, Lucas J. Fern\'andez-Alc\'azar$^{1,2}$, Zin Lin$^{3}$,  Tsampikos Kottos$^{1}$\\}

\affiliation{
$^{1}$Wave Transport in Complex Systems Lab, Physics Department,\\
Wesleyan University, Middletown, CT-06459, USA\\
$^{2}$Institute for Modeling and Innovative Technology
(CONICET - UNNE) and \\
Natural \& Exact Science Faculty, Northeastern University\\ 
Corrientes, W3404AAS, Argentina\\
$^{3}$Bradley Department of Electrical and Computer Engineering, Virginia Tech, Blacksburg, VA-24060, USA}


\date{\today}
\begin{abstract}
Adjoint methods provide a powerful route for gradient-based optimization, but their physical implementation is obstructed in generic nonlinear systems because the adjoint dynamics requires backward-time evolution, Jacobian transposition, and terminal-value constraints. Here we show that nonlinear parity-time ($\mathcal{PT}$)-symmetric systems overcome this obstruction. Using a class of nonlinear non-Hermitian resonator networks, we establish symmetry relations that map the formal adjoint dynamics onto experimentally accessible forward-time evolutions supplemented by controlled injections. This construction enables exact in-situ evaluation of adjoint gradients without requiring explicit backward propagation or matrix transposition. We demonstrate the approach in nonlinear $\mathcal{PT}$-symmetric resonator chains, where the resulting optimization protocol autonomously discovers parameter configurations that realize prescribed spatio-temporal functionalities, including uniform energy redistribution and targeted wave transport at predefined time windows. Our results identify $\mathcal{PT}$ symmetry as a resource for implementing computational sensitivities within physical systems and establish a route toward self-optimizing nonlinear machines.
\end{abstract}
\maketitle

\section{Introduction}
When a wave scatters again and again through a complex nonlinear medium, the interference of its countless multiply-scattered paths produces an input-output map of extraordinary richness. Long regarded as disorder to be suppressed, this complexity is increasingly recognized as a resource: the same tangled, high-dimensional response that resists analytic description can be repurposed to filter, classify, remember, and transform signals directly in the physical layer, giving rise to a broad family of wave-based and physical computing schemes \cite{psaltis1990holography, shen2017deep, lin2018all, Wetzstein2020DeepOptics, Shastri2021PhotonicsAI, wright2022, Fei2024, Fleury2025}. Nonlinearity is the essential ingredient.
It is what lifts a scattering network from a fixed linear transformation into a substrate capable of genuinely expressive computation.

What such substrates have conspicuously lacked is the capacity to improve themselves. In machine learning, that capacity rests entirely on the gradient. In a physical substrate, the same gradient is supplied by the adjoint method: backpropagation through the substrate's own equations of motion. A single backward pass returns the sensitivity of a scalar objective to every parameter at once, at a cost independent of their number \cite{Hughes2018NonlinearAdjoint, molesky2018inverse}. It is this single primitive that makes optimization over millions of parameters feasible. 
Recently, the same primitive has begun to migrate out of the digital computer and into the physical device itself, with schemes that evaluate adjoint gradients in situ, letting a system measure its own sensitivities~\cite{PSHPBWMMA2023, GWLK2025, DMW2025, guillamon2026insituadjointwavecontrol}. 
Related approaches have also explored physical backpropagation through nonlinear optical systems and alternative backpropagation-free training strategies \cite{Guo2021NonlinearBackprop,Spall2025OpticalBackprop,Momeni2023}.
The prospect is tantalizing: a medium that computes its own gradients could tune itself toward a goal, adapt when its task or environment changes, and operate without any external model of its own dynamics. Self-optimization of this kind is the seed from which adaptability, autonomy, and ultimately a physically embodied form of intelligence might grow.

This prospect, however, collides with a fundamental obstruction, and it does so exactly where the physics is most interesting. The adjoint field is governed by an equation that must (i) evolve backward in time, (ii) be generated by the transpose of the system's Jacobian, and (iii) be fixed by a terminal rather than an initial condition. In a linear, reciprocal, energy-conserving medium these three demands can occasionally be met through time-reversal or reciprocity~\cite{PSHPBWMMA2023, GWLK2025, hughes2018, LM23}, which is why earlier in-situ gradient schemes have relied on precisely those properties, the very ones that make a substrate computationally bland. 
Alternative approaches have exploited spatial symmetry and reciprocity to eliminate explicit backward propagation in physical training protocols~\cite{Xue2024FullyForward}.
But in the nonlinear, non-Hermitian systems that are the richest computational substrates, those with gain, loss, and asymmetric couplings that explicitly break microscopic reversibility, none of the three corresponds to any realizable physical process. The transpose Jacobian, in particular, bears no simple relation to anything the forward dynamics can produce. The adjoint remains mathematically exact yet physically inert, and the dream of in-situ self-optimization stalls at the nonlinear frontier where it would matter most.

A resolution comes from an idea of an entirely different lineage. Symmetry is the most powerful organizing principle in physics: through Noether's theorem it fixes the conservation laws, and through gauge principles it dictates the very form of the equations of motion. Symmetry is generative, not merely restrictive; it can determine what the dynamics are allowed to be. Parity-time (PT) symmetry, the balanced arrangement of gain and loss that has reshaped non-Hermitian physics over the past two decades~\cite{BenderBoettcher1998,Bender2007,Guo2009,ruter2010observation, schindler2012symmetric,Peng2014,Fleury2015,Feng2017, ElGanainy2018, lin2011unidirectional, el2018non,Ozdemir2019, Kafesaki2019,Assawaworrarit2017,Bai2023}, is almost always invoked for its spectral consequences, such as real eigenvalues and exceptional points. Here we ask a different question: whether symmetry can be made to act on computation rather than on spectra, converting an abstract algorithmic operation into a physically admissible one.

The interplay between the adjoint, an optimization construct, and physical symmetry, a property of dynamical laws, has, to our knowledge, never been examined. We show that the two are, in a precise sense, made for each other. A class of nonlinear PT-symmetric resonator networks possesses exactly the structural identities needed to meet each of the adjoint's three demands in turn: PT symmetry converts backward-time evolution into ordinary forward-time evolution, while a fixed phase-space involution converts the transpose Jacobian into the untransposed one, in both cases up to a known injection that can be supplied externally from the already-measured forward trajectory. As a result, the full adjoint computation, transpose Jacobian and terminal condition included, is emulated by forward-time evolution of the physical system alone. The construction is exact in the limit of vanishing injection strength and, decisively, holds for fully nonlinear dynamics rather than a weakly nonlinear approximation. Throughout, the optimization is constrained to preserve the PT symmetry that makes it physical, so the machine improves itself without ever leaving the class of systems in which self-improvement is possible. Using this protocol, a nonlinear PT-symmetric chain autonomously tunes its own couplings, nonlinearities, and gain-loss profile to realize prescribed spatio-temporal functionalities, from uniform energy redistribution to targeted wave transport within predefined time windows. Beyond the specific protocol, these results identify symmetry as an operational resource for physical computation and mark a concrete step toward self-optimizing nonlinear machines.

\begin{figure*}
\begin{center}
 \includegraphics[width=1\textwidth]{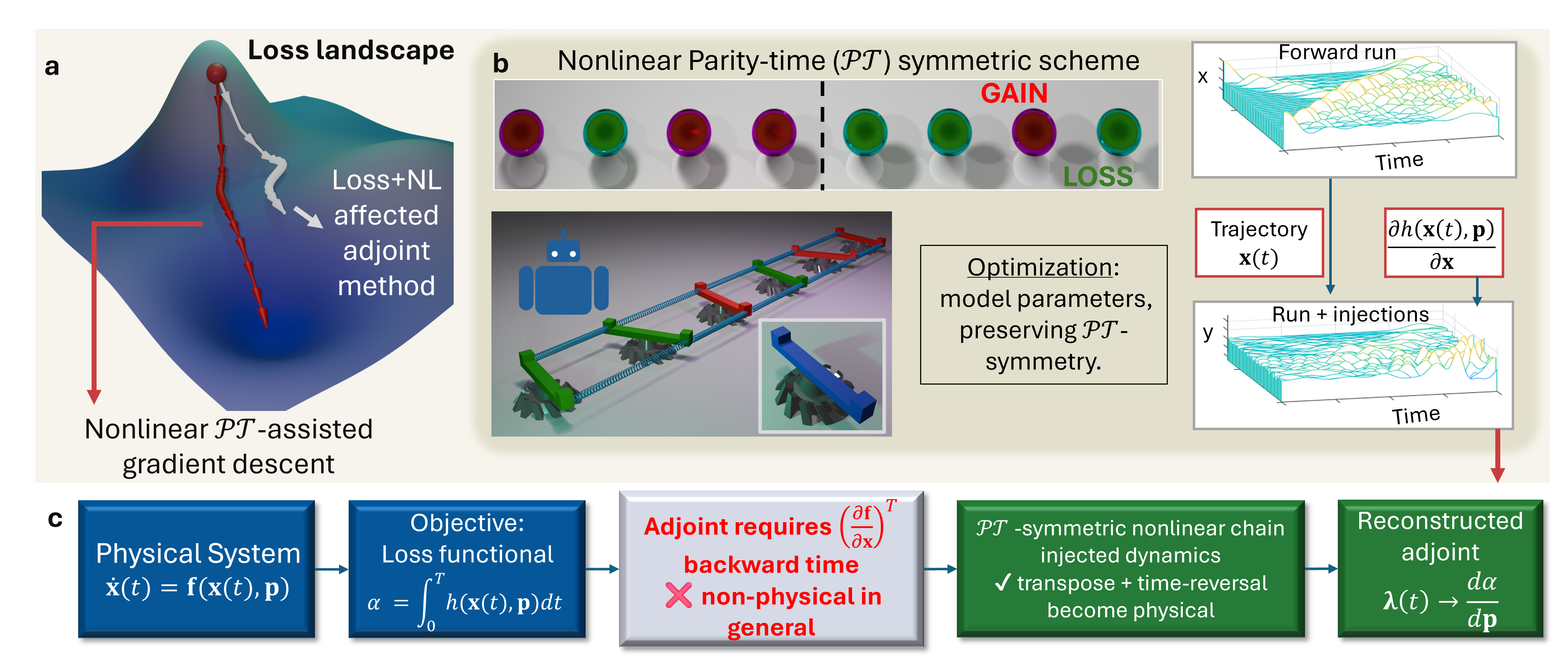}
  \end{center}
  \caption{{\bf PT-assisted physical adjoint optimization protocol for nonlinear nonreciprocal/ lossy systems.} Conceptual overview of the ${\cal PT}$-assisted in-situ adjoint method. A nonlinear physical system is first evolved forward in time to evaluate the objective functional. While the adjoint problem formally requires backward-time evolution and the transpose Jacobian, these operations are not physically accessible in generic non-reciprocal or lossy systems. By exploiting ${\cal PT}$-symmetry and controlled injections in a ${\cal PT}$-nonlinear resonator chain, the adjoint dynamics can be mapped onto a forward, experimentally accessible evolution, enabling in-situ evaluation of the gradient with minimal processing.}
  \label{fig1}
\end{figure*}

\section{Conceptual problem: adjoint sensitivity beyond time-reversal symmetry in a nonlinear system.}

Adjoint methods provide an efficient framework for computing the sensitivities of a scalar objective function with respect to a large number of system parameters~\cite{Cea1986Adjoint,molesky2018inverse,Hughes2018NonlinearAdjoint}. Formally, for a nonlinear dynamical system $\dot{\mathbf{x}}(t)=\mathbf{f}(\mathbf{x}(t),\mathbf{p})$ and an objective functional 
\begin{equation}
    \alpha(\mathbf{p})=\int_0^T h(\mathbf{x}(t),\mathbf{p}) dt,
\end{equation}
the gradient $d\alpha/d\mathbf{p}$ can be expressed in terms of an adjoint field $\mathbf{\lambda}(t)$ that satisfies an equation involving the transpose of the Jacobian of the vector field, $\mathbf{ \dot \lambda}+\left(\frac{\partial \mathbf{f}}{\partial \mathbf{x}}\right)^T \mathbf{\lambda}=-\left(\frac{\partial h}{\partial \mathbf{x}}\right)^T$ with terminal condition $\mathbf{\lambda}(T)=0$ at the final time $T$~\cite{TDadjoint}.

From a mathematical point of view, this construction is entirely general and does not rely on specific symmetries of the underlying system. From a physical standpoint, however, the adjoint formulation poses a fundamental difficulty. The adjoint problem typically requires (i) backward-time evolution, (ii) access to the transpose Jacobian $(\partial \mathbf{f}/\partial \mathbf{x})^T$, and (iii) the imposition of a terminal condition. None of these operations correspond, in general, to a realizable physical process in nonlinear systems that lack time-reversal symmetry. This obstruction is particularly severe in non-Hermitian and non-reciprocal platforms, where gain, loss, and asymmetric couplings explicitly break microscopic reversibility.

This limitation is not merely technical, but reflects a structural mismatch between the formal adjoint construction and the set of operations accessible to a physical system. Bridging this gap requires identifying structural conditions under which backward-time evolution and Jacobian transposition can be mapped onto physically accessible transformations.

\section{Conceptual strategy: ${\cal PT}-$symmetry as a structural enabler}

The central idea of this work is that parity–time (${\cal PT}$) symmetry provides precisely the structural ingredient needed to overcome the aforementioned obstruction. Rather than treating ${\cal PT}$-symmetry as a spectral property, we exploit it here as an operational tool that enables nontrivial mappings between forward and backward dynamics, as well as between a Jacobian and its transpose.

We focus on a class of nonlinear ${\cal PT}$-symmetric resonator chains, in which gain and loss are distributed antisymmetrically with respect to the system center, while couplings and nonlinearities are parity symmetric (we exclude asymmetric couplings). Although our formulation is general and can be applied to a variety of situations, we will adopt, for simplicity, the language of electromagnetic fields confined to resonators as in Fig. \ref{fig1}. In such a case, the field dynamics in a 1D chain of $2N$ resonators is determined by a temporal coupled mode theory 
$i\frac{d\mathbf{\Psi} }{dt}=\mathbf{H}_{PT}\,\mathbf{\Psi}+i\mathbf{W} \mathbf{S}_0^+$
where $\mathbf{\Psi}=(\psi_1,\cdots,\psi_{2N})^T$, being $\psi_j$ the field amplitude at resonator $j=1,\cdots,2N$. The last term represents the external input from sources. 
The Hamiltonian $\mathbf{H}_{PT}$ that dictates the field evolution has diagonal elements $H_{jj}= i\gamma_j+g_j(|\psi_j|^2)$, where $\gamma_j>0 \quad (<0)$ represents the loss (gain) rate, and the functions $g_j(|\psi_j|^2)\in {\bb R}$ correspond to the nonlinear frequency shifts. The off-diagonal elements determine the couplings through the chain $H_{j,j+1}=H_{j+1,j}=\kappa_{j}$.
When expressed in real ``position-momentum" coordinates, $\mathbf{x}\equiv (q_1, p_1, \cdots q_{2N},p_{2N})$ with $q_j=\Re(\psi_j)$ and $p_j=\Im(\psi_j)$, the resulting dynamics $\mathbf{\dot x}=\mathbf{f}(\mathbf{x})$ defines a non-Hermitian, nonlinear vector field $\mathbf{f}$ that nonetheless satisfies a set of discrete symmetry relations involving the action of parity (${\cal P}$), time reversal (${\cal T}$), and their combination.

Crucially, these symmetries will allow us to overcome the difficulties that arise in the adjoint problem implementation by introducing a mapping between 
a forward-time trajectory and its backward-time counterpart. At the same time, for this specific class of systems, the Jacobian of the nonlinear dynamics exhibits a nontrivial relation between transposition and time reversal, up to a symmetry-induced correction term that can itself be implemented as a controlled injection.
To be specific, the following symmetry identities hold 
\begin{enumerate}
\item ${\cal PT}$-symmetry of the vector field, ${\cal PT}\mathbf{f}({\cal PT} \mathbf{x})=-\mathbf{f}(\mathbf{x})$;
\item ${\cal PT}$-symmetry of the trajectory $\mathbf{x}(t)$, ${\cal PT}\mathbf{x}(t)=\mathbf{x}(-t)$, provided that the initial condition is ${\cal PT}$-symmetric, ${\cal PT}\mathbf{x_0}=\mathbf{x_0}\equiv \mathbf{x}(0)$;
\item Parity-reversal via $\mathbf{\Gamma}$-injection: $\mathbf{f}(\mathbf{x})+\mathbf{\Gamma} \mathbf{x} = {\cal P} \mathbf{f} ({\cal P} \mathbf{x})$.
\item Transpose and Time-reverse via a transformation matrix $\mathbf{\mathbf{\Theta}}$: $\mathbf{\Theta} \frac{\partial \mathbf{f}}{\partial \mathbf{x}} [\mathbf{x}(t)]\mathbf{\Theta} = \left( \frac{\partial \mathbf{f}}{\partial \mathbf{x}}\right)^T[\mathbf{x}(-t)] +\mathbf{\Gamma}$, which hold for a ${\cal PT}$-symmetric initial condition, ${\cal PT}\mathbf{x_0}=\mathbf{x_0}$, $\mathbf{\Theta}_{j,k}\equiv \delta_{j,4N-k+1}$, and $\mathbf{\Gamma}\equiv -2 {\rm diag}(\gamma_1,\gamma_1, \cdots, \gamma_{2N}, \gamma_{2N})$, where each gain/loss coefficient $\gamma_j$ acts identically on the conjugate pair $(q_j,p_j)$.
\end{enumerate}
These identities are non-generic and rely crucially on the specific ${\cal PT}$-symmetric structure of the nonlinear resonator chain.
Taken together, these properties imply that the formal operations appearing in the adjoint equation—backward-time evolution and Jacobian transposition—can be emulated by forward-time evolution of a physical system with suitably engineered injections. Notably, this construction does not rely on linearity or weak nonlinearity, but holds for fully nonlinear dynamics, provided the ${\cal PT}$-symmetry constraints are satisfied.

From a conceptual perspective, the role of ${\cal PT}$-symmetry here is not to simplify the adjoint equation, but to re-encode it into a physically realizable dynamical process. This re-encoding is the key step that allows adjoint-based sensitivity analysis to be performed in situ, directly within the physical platform under investigation.
For a pictorial description of the above discussion, see Extended Figure \ref{fig:ExtFig1}.

\section{PT-assisted in-situ adjoint protocol}

The conceptual workflow enabled by this symmetry structure is summarized in Fig. \ref{fig1} and consists of three main stages 
(see also Extended Figure \ref{fig:ExtFig1}).

First, the physical system is evolved forward in time from a PT-symmetric initial condition, $\mathbf{\dot x}=\mathbf{f}(\mathbf{x})$ with $\mathbf{x_0}={\cal PT}\mathbf{x_0}$. This forward evolution generates the trajectory $\mathbf{x}(t)$ and allows direct evaluation of the objective functional. No modification of the physical dynamics is required at this stage.

Second, an auxiliary evolution is performed in the {\bf same} physical platform, but with controlled injections that depend on the previously recorded forward trajectory. Specifically, an evolution $\mathbf{\dot y}=\mathbf{f}(\mathbf{y}) + \mathbf{\Gamma} \mathcal{T} \mathbf{x}(-t) +\epsilon {\cal P} \mathbf{\Theta} \left[ \frac{\partial h}{\partial \mathbf{x}}(\mathbf{x}(-t)) \right]^T$ from $-T$ to $0$ is performed with initial condition $\mathbf{y}(-T)={\cal T}\mathbf{x}(T)$. Here, $\epsilon$ corresponds to a small dimensionless number controlling the injection strength. Owing to the ${\cal PT}$-symmetry relations discussed above, this evolution with injections effectively implements a time-reversed and transposed version of the original dynamics, up to known and controllable correction terms. Importantly, this stage involves only forward-time integration of a physical system, despite encoding the information content of the adjoint equation.

Third, the adjoint field is reconstructed from the auxiliary trajectory by applying symmetry operations that are fixed and independent of the system parameters, 
$\mathbf{\lambda}(t)=\frac{1}{\epsilon} \mathbf{\Theta} \left( {\cal P} \mathbf{y}(-t) -
{\cal{P} \cal{T}} \mathbf{x}(t) \right)$. Once the adjoint field is obtained, the gradient of the objective with respect to the system parameters follows directly from a time integral involving the forward trajectory and the reconstructed adjoint field $\frac{d\alpha}{d\mathbf{p}}=\int_0^T \left( \frac{\partial h}{\partial \mathbf{p}}(\mathbf{x}(t)) + \mathbf{\lambda}^T(t) \frac{\partial \mathbf{f}}{\partial \mathbf{p}}(\mathbf{x}(t)) \right)$.

This protocol yields the exact adjoint gradient in the limit of vanishing injection strength $\epsilon$, while remaining entirely within the space of physically realizable operations. In contrast to conventional numerical adjoint methods, no explicit backward integration, matrix transposition, or terminal constraints need to be imposed externally.

Beyond its immediate application to gradient-based optimization, this framework establishes a broader principle: discrete symmetries can be leveraged not only to shape spectra or dynamics, but also to enable otherwise inaccessible computational primitives within \textit{in-situ} physical protocols. In this sense, ${\cal PT}$-symmetry acts as a bridge between abstract adjoint calculations and physically implementable dynamics.
More broadly, this suggests that symmetry-engineered nonlinear platforms may serve as autonomous computational substrates for optimization and learning tasks.

\begin{figure}
\begin{center}
 \includegraphics[width=0.45\textwidth]{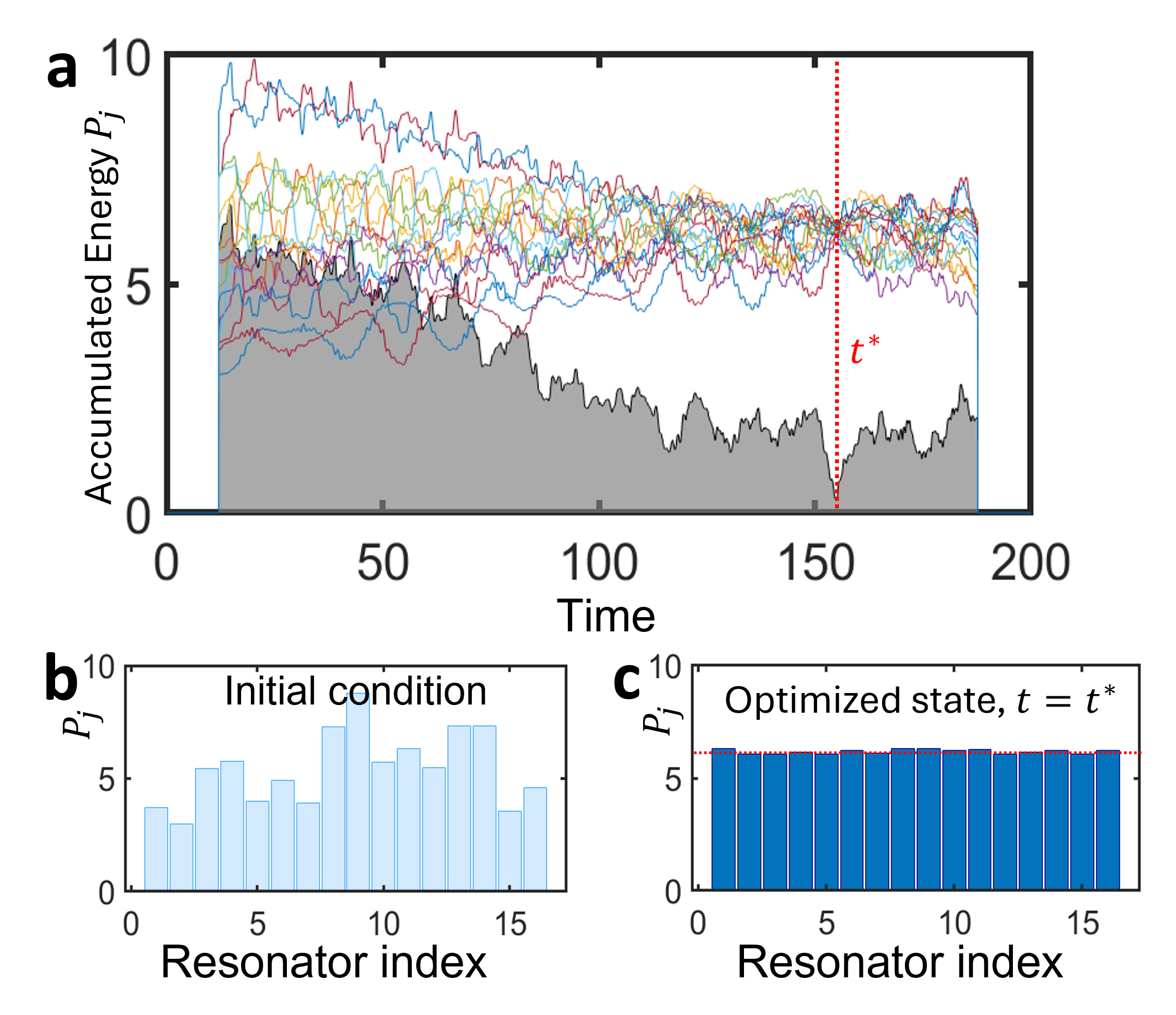}
  \end{center}
\caption{
\textbf{a,} Time-resolved accumulated energy 
$P_j(t;\mathbf{p})
=
\int_0^T dt'\,
w_t(t')\,
\left|\psi_j(t';\mathbf{p})\right|^2$ for all $2N=16$ resonators (coloured lines). 
The gray shaded region represents the time-windowed spread
$\Delta P(t;\mathbf{p}^\star)
=
\max_{1\leq j\leq N}P_j(t;\mathbf{p}^\star)
-
\min_{1\leq j\leq N}P_j(t;\mathbf{p}^\star)$.
The optimization minimizes this spread at the target time $t^\ast=155$
(red dashed line). 
\textbf{b,} Initial accumulated energy showing a non-uniform distribution. 
\textbf{c,} Optimized state at $t^*$ exhibiting uniform energy distribution 
(red dashed line is a guide to the eye).
}
\label{fig2}
\end{figure}

\section{Application cases}

\subsection{Uniform energy redistribution at a specific time-window}

To illustrate our method, we consider a one-dimensional chain of $2N=16$ coupled resonators. Each resonator exhibits a Kerr-type nonlinearity of strength $\chi_j$, such that the nonlinear frequency shift results $g_j(|\psi_j|^2)\equiv \omega_j+\chi_j |\psi_j|^2$. Each $j$-th resonator is coupled to its nearest neighbor $j+1$ with coupling constant $\kappa_j$. All resonators are passive (no loss or gain) except for a central non-Hermitian dimer defect: the resonator $j=8$ is endowed with linear gain $+\gamma$, while the resonator $j=9$ experiences an equal amount of linear loss $-\gamma$. 

The parameters $\{\kappa_j,\chi_j,\gamma_j\}$ are collected into the design vector $\mathbf{p}$ and optimized using the adjoint-based protocol described above. To quantify the energy accumulated at each resonator over a temporal window of duration $\tau$ centered at time $t$, we define
\begin{equation}
P_j(t;\mathbf{p})
=
\int_0^T dt'\,
w_t(t')\,
\left|\psi_j(t';\mathbf{p})\right|^2,
\label{eq:windowed_energy}
\end{equation}
where
\begin{equation}
w_t(t')
=
\begin{cases}
1, & t'\in[t-\tau/2,t+\tau/2],\\
0, & \text{otherwise}.
\end{cases}
\label{eq:window_function}
\end{equation}
The optimization objective is the spatial spread of these window-integrated energies at the prescribed target time $t^\ast$,
\begin{equation}
\alpha(\mathbf{p})
=
\Delta P(t^\ast;\mathbf{p})
\equiv
\max_{1\leq j\leq N} P_j(t^\ast;\mathbf{p})
-
\min_{1\leq j\leq N} P_j(t^\ast;\mathbf{p}).
\label{eq:uniformity_objective}
\end{equation}
Minimizing $\alpha(\mathbf{p})$ penalizes differences among the energies accumulated by the individual resonators over the target window and therefore promotes a spatially uniform distribution of window-integrated energy across the chain. Here, the target window is centered at $t^\ast\simeq155$, with $\tau=15$ and total evolution time $T=200$.

For visualization, Fig.~\ref{fig2}(a) shows $P_j(t;\mathbf{p}^\star)$ as a function of the window-center time $t$ for all resonators in the optimized configuration $\mathbf{p}^\star$. The corresponding time-dependent spread,
\begin{equation}
\Delta P(t;\mathbf{p}^\star)
=
\max_{1\leq j\leq N}P_j(t;\mathbf{p}^\star)
-
\min_{1\leq j\leq N}P_j(t;\mathbf{p}^\star),
\label{eq:time_resolved_spread}
\end{equation}
is indicated by the gray shaded region and directly measures the nonuniformity of the accumulated-energy distribution. The spread reaches a pronounced minimum near the prescribed target time $t^\ast$, showing that the resonators accumulate nearly equal amounts of energy within the observation window centered at $t^\ast$.

The system is initialized in a strongly non-uniform state, where the energy is unevenly distributed along the chain; see Fig. \ref{fig2}(b). As the optimization proceeds and the parameters approach their optimal values, the dynamical evolution reshapes the energy flow such that the spatial distribution progressively homogenizes. This behavior is most clearly visualized by inspecting the instantaneous time-windowed energy at the target time: while the initial condition exhibits large fluctuations across sites, the optimized system displays an almost perfectly uniform energy distribution across all sixteen resonators; see  Fig. \ref{fig2}(c). 
During the optimization procedure, the relative energy spread $\alpha(\mathbf{p})/\langle P_j \rangle$
is reduced from $106\%$ in the initial state to $4\%$ after optimization.

\subsection{Optimized spatio-temporal wave transport}

\begin{figure*}
\begin{center}
\includegraphics[width=1\textwidth]{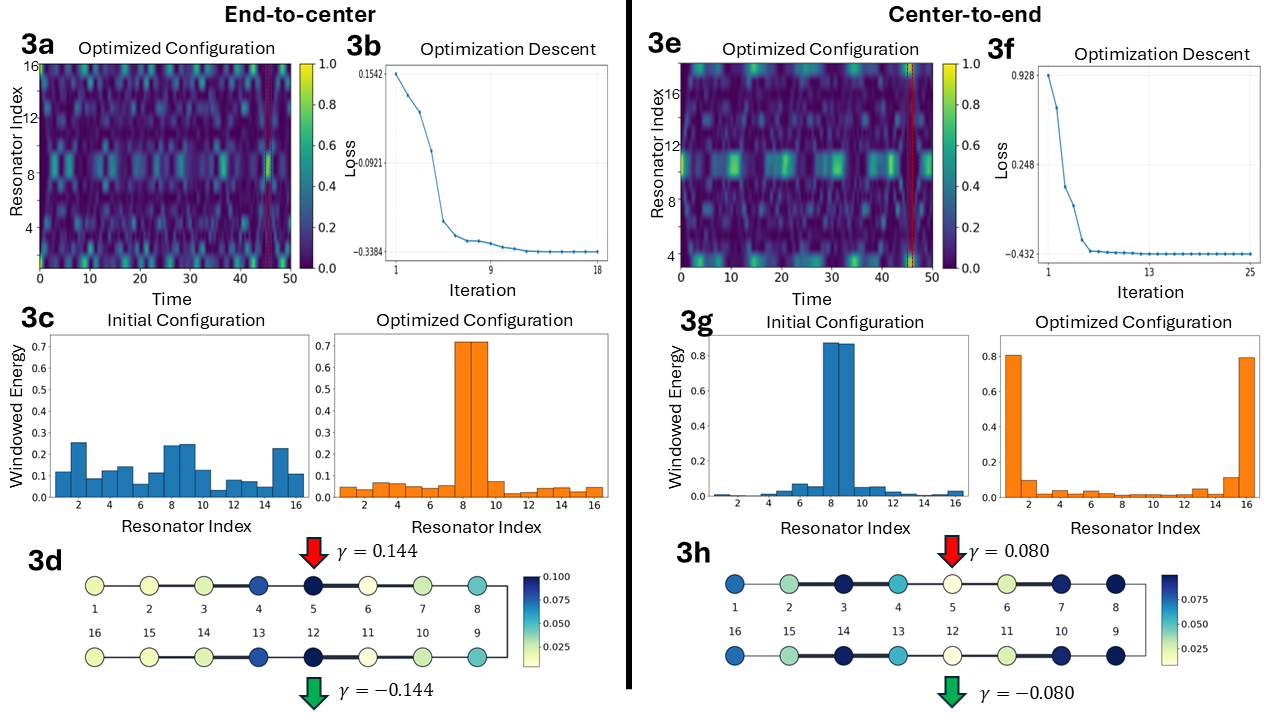} 
  \end{center}
 \caption{
{\bf Temporal heatmaps and energy distributions for the nonlinear transport tasks.} Figure 3 shows the end-to-center task, where energy initialized at the ends is concentrated near the middle of the chain. Figure 4 shows the center-to-end task, where energy initialized near the middle is transported toward and concentrated at the ends of the chain. 
\textbf{3 a,e.} Temporal evolution of the resonator intensities for the optimized configuration. The shaded red region indicates the target time window.
\textbf{3 b,f.} Descent of the optimization objective during the gradient descent trial that produced the optimized configuration.
\textbf{3 c,g.} Energy distribution within the target time window for the initial and optimized configurations.
\textbf{3 d,h.} Schematic of the optimized configuration. The colormap represents the nonlinearity parameter $\chi$, the line thickness between neighboring resonators represents the coupling strength $\kappa$, and the red/green markers indicate the gain/loss coefficients $\gamma$. 
We have used $T=50$, $t^\ast=45$, and $\tau=0.5$. }
\label{fig3}
\end{figure*}

We next employ the adjoint-based optimization protocol to perform dramatically different energy transport tasks, depending on the target objective function.
The setup is again the nonlinear non-Hermitian one-dimensional resonator chain described in the previous section, 
and the optimization variables are the coupling constants $\kappa_j$, the nonlinear coefficients 
$\chi_j$, and the gain/loss strengths $\gamma_j$. 
In this case, the non-Hermitian elements are symmetrically displaced from the center: resonator $j=5$ is endowed with linear gain $+\gamma$, while resonator $j=12$ experiences an equal amount of linear loss $-\gamma$. All other resonators remain passive. 
The objective is to engineer the network parameters such that the system performs prescribed transport operations between different spatial regions of the chain.

We first consider an end-to-center transport task, where the energy is initially localized near the end of the chain, the optimized dynamics should concentrate the energy near the center sites $j=8,9$ at the target time window $[t^\ast-\tau/2, t^\ast+\tau/2]$.  
To quantify the performance of the process, we define the following optimization objective,
which uses smooth approximations of the maximum and minimum functions to avoid non-differentiability
\begin{equation}
\begin{aligned}
\alpha_{\mathrm{center}}(\mathbf p)
&=
\operatorname{smax}_{\nu}
\left\{
P_j(t^\ast;\mathbf p)
\right\}_{j\in\mathcal A}
\\
&\quad -
\operatorname{smin}_{\nu}
\left\{
P_j(t^\ast;\mathbf p)
\right\}_{j\in\mathcal A^C},
\end{aligned}
\label{eq:center_objective}
\end{equation}
with $\mathcal{A}=\{1,\ldots,16\}-\{8,9\}$ and $\mathcal{A}^C=\{8,9\}$. Specifically, for a set of real values \(\{x_j\}_{j\in S}\), we define
\[
\operatorname{smax}_{\nu}\{x_j\}_{j\in S}
=
\nu \log\left(\sum_{j\in S} e^{x_j/\nu}\right),
\]
\[
\operatorname{smin}_{\nu}\{x_j\}_{j\in S}
=
-\nu \log\left(\sum_{j\in S} e^{-x_j/\nu}\right),
\]
with $\nu=0.1$.
Here, the dynamics is initialized from an end-localized state. Minimizing this objective favors configurations in which the two central resonators simultaneously possess higher energy than any other site during the target time window.

The results of such optimization scheme are shown in Fig. \ref{fig3}a-d. For the unoptimized initial configuration, the energy is not effectively concentrated at the central sites at the prescribed target time. In contrast, 
the optimized dynamics successfully redirects the initially edge-localized excitation toward the center of the chain. This behavior is clearly visible in the spatio-temporal maps of Fig. \ref{fig3}a, where the intensity becomes concentrated around resonators $j=8$ and $j=9$ within the target time window. The optimization converges rapidly, reaching a nearly stationary objective value after fewer than 20 iterations (Fig. \ref{fig3}b).

This improvement is also evident from the bar plots in Fig. \ref{fig3}c. For the initial configuration, the time-window-averaged energy distribution remains uneven and does not display clear localization at the target sites. In contrast, after optimization, the energy becomes strongly concentrated at the two central resonators. After optimization, approximately $70\%$ of the total energy contained in the chain is concentrated in the two target resonators, demonstrating highly efficient transport and localization.

Finally, Fig. \ref{fig3}(d) shows the optimized system parameters. Interestingly, the optimized solution develops a structured landscape of couplings and nonlinearities around the target region, together with a finite gain-loss contrast, indicating that the transport functionality emerges from a cooperative interplay between interferences  and non-Hermitian effects.

We now turn to the complementary transport problem, in which energy is initially concentrated near the center of the chain and the optimized dynamics should concentrate the energy near the two end sites $j=1,16$ at the same target time window.
Similarly, for the center-to-end task, we define
\begin{equation}
\begin{aligned}
\alpha_{\mathrm{end}}(\mathbf p)
&=
\operatorname{smax}_{\nu}
\left\{
P_j(t^\ast;\mathbf p)
\right\}_{j\in\mathcal B}
\\
&\quad -
\operatorname{smin}_{\nu}
\left\{
P_j(t^\ast;\mathbf p)
\right\}_{j\in\mathcal B^C}.
\end{aligned}
\label{eq:end_objective}
\end{equation}
with $\mathcal{B}=\{1,\ldots,16\} - \{1,16\}$ and $\mathcal{B}^C=\{1,16\}$.
Minimizing this objective favors simultaneous energy concentration at both ends of the chain.

The optimized dynamics displays a qualitatively different behavior from the previous example. Instead of funneling energy toward a central region, the optimization discovers a configuration that efficiently splits the initially localized excitation and routes it toward the two ends of the chain.
As shown in Fig. \ref{fig3}e, the energy remains localized near the central resonators during the early stages of the evolution and subsequently propagates outward, reaching the target edge sites within the prescribed time window.
Remarkably, such optimization process also takes about 20 iteration steps (Fig. \ref{fig3}f), indicating the robustness of the method.

Fig. \ref{fig3}g further confirms the performance of the optimization scheme.
Whereas the initial configuration is unable to efficiently route energy toward the target sites, the optimized dynamics redirects approximately $77\%$ of the total energy to the two edge resonators within the prescribed time window, demonstrating highly efficient center-to-end transport.
Figure \ref{fig3}h displays
the corresponding optimized parameter landscape, revealing a markedly
different arrangement of couplings and nonlinearities from that found
for the end-to-center task.

The comparison between the protocols in Figs. \ref{fig3} highlights a central feature of the proposed optimization framework. The underlying physical platform is identical in both cases; only the objective function is modified. Nevertheless, the optimization results in two markedly different parameter landscapes that implement opposite transport operations, demonstrating that the $\cal{PT}$-assisted adjoint protocol does not optimize a specific physical mechanism, but rather identifies system configurations capable of realizing a desired spatio-temporal  functionality.

\section{Conclusion}

We have shown that nonlinear $\mathcal{PT}$ symmetry can be exploited as a structural mechanism for realizing adjoint dynamics within a physical system. By combining symmetry-induced mappings with controlled injections, the formal operations required by the adjoint problem—backward-time evolution, Jacobian transposition, and terminal constraints—are transformed into experimentally accessible forward-time processes. This enables exact in-situ gradient evaluation in nonlinear non-Hermitian platforms and removes a fundamental obstacle to in-situ optimization in nonlinear physical systems.

Beyond the specific resonator platform considered here, our work suggests a broader role for symmetry in physical computation. Rather than serving only to constrain dynamics or shape spectral properties, nonlinear $\mathcal{PT}$ symmetry functions here as a computational resource, converting the abstract operations required for adjoint optimization into experimentally realizable dynamical processes. If nonlinear wave systems are to become adaptive computational substrates, they must be able not only to perform a task, but also to evaluate the information required to improve their own performance. The framework developed here provides a concrete route toward that objective, demonstrating how a physical system can compute its own optimization gradients while remaining entirely within its native dynamics. More generally, our results suggest that symmetry may determine not only the dynamics a physical system can realize, but also the computational primitives it can implement~\cite{Vadlamani2020Lagrange,Momeni2023}, opening a path toward a new generation of self-optimizing nonlinear physical machines.

\begin{acknowledgements}
Z.Li, L.J.F.-A., and T.K. acknowledge partial support from Simons Foundation  SFI-MPS-EWP-00008530-08. 
L.J.F.-A. acknowledges the hospitality of Wesleyan University, where part of this work was
developed, and partial support from CONICET and (ex)MINCyT grant number CONVE-2023-10189190-FFFLASH.
Z.Lin is supported by the U.S. Army Research Office (award numbers W911NF2410390 and W911NF2510113).
T.K. acknowledges partial support from MURI ONR-N000142412548. 
\end{acknowledgements}

\bibliographystyle{unsrt} 
\bibliography{refs}

\section{Methods}
\subsection{Adjoint sensitivity analysis and conceptual obstruction}
\label{sec:adjoint_obstruction}

We begin by recalling the standard time-domain adjoint formulation for sensitivity analysis in nonlinear dynamical systems \cite{TDadjoint}, emphasizing the steps that will later reveal the fundamental physical obstruction.

We consider a nonlinear dynamical system 
\begin{equation}
\dot{\mathbf{x}}(t)=\mathbf{f}\bigl(\mathbf{x}(t),\mathbf{p}\bigr), \qquad \mathbf{x}(0)=\mathbf{x}_0,
\label{eq:forward_dynamics}
\end{equation}
where $\mathbf{x}(t)\in\mathbb{R}^n$ is the state vector and $\mathbf{p}\in\mathbb{R}^m$ denotes a set of parameters. The loss functional is assumed to be of the form
\begin{equation}
\alpha(\mathbf{p})=\int_0^T h\bigl(\mathbf{x}(t),\mathbf{p}\bigr)\,dt,
\label{eq:objective}
\end{equation}
with $T>0$ fixed and $h\bigl(\mathbf{x}(t),\mathbf{p}\bigr)$ some smooth function of $\mathbf{x}(t)$ and $\mathbf{p}$ (such that the derivatives $\partial h/\partial \mathbf{x}$  and $\partial h/\partial \mathbf{p}$ exist and are bounded). 

Our goal is to compute the gradient $d\alpha/d\mathbf{p}$ without explicitly integrating $n$ auxiliary equations for every one of the $m$ parameters. These problems are efficiently addressed by the adjoint method. 

{\it Lagrangian formulation.-}
The adjoint method proceeds by introducing the dynamics Eq. \eqref{eq:forward_dynamics} through a Lagrange multiplier -- the adjoint field $\boldsymbol{\lambda}(t)\in\mathbb{R}^n$ -- into the augmented functional
\begin{equation}
\mathcal{L}(\mathbf{x},\mathbf{p},\boldsymbol{\lambda})
=
\int_0^T
\left[
h\bigl(\mathbf{x}(t),\mathbf{p}\bigr)
-
\boldsymbol{\lambda}^T(t)\left(\dot{\mathbf{x}}(t)-\mathbf{f}\bigl(\mathbf{x}(t),\mathbf{p}\bigr)\right)
\right]dt.
\label{eq:lagrangian}
\end{equation}
Since $\dot{\mathbf{x}}(t)-\mathbf{f}(\mathbf{x}(t),\mathbf{p})=0$, the total derivative satisfies
\begin{equation}
\frac{d\mathcal{L}}{d\mathbf{p}}=\frac{d\alpha}{d\mathbf{p}}.
\end{equation}
Taking the total derivative of \eqref{eq:lagrangian} with respect to $\mathbf{p}$ yields
\begin{eqnarray}
\frac{d\mathcal{L}}{d\mathbf{p}}
&=&
\int_0^T
\left(
\frac{\partial h}{\partial \mathbf{x}}\frac{\partial \mathbf{x}}{\partial \mathbf{p}}
+
\frac{\partial h}{\partial \mathbf{p}}
\right)dt  \notag \\
&-&
\int_0^T
\boldsymbol{\lambda}^T
\left(
\frac{d}{dt}\frac{\partial \mathbf{x}}{\partial \mathbf{p}}
-
\frac{\partial \mathbf{f}}{\partial \mathbf{x}}\frac{\partial \mathbf{x}}{\partial \mathbf{p}}
-
\frac{\partial \mathbf{f}}{\partial \mathbf{p}}
\right)dt .
\label{eq:dLdp}
\end{eqnarray}
Integrating by parts the term involving $\frac{d}{dt}\frac{\partial \mathbf{x}}{\partial \mathbf{p}}$, we obtain
\begin{eqnarray}
\frac{d\mathcal{L}}{d\mathbf{p}}
&=&
\int_0^T
\frac{\partial h}{\partial \mathbf{p}}\,dt
+
\int_0^T
\boldsymbol{\lambda}^T\frac{\partial \mathbf{f}}{\partial \mathbf{p}}\,dt
-
\Bigl[\boldsymbol{\lambda}^T \frac{\partial \mathbf{x}}{\partial \mathbf{p}}\Bigr]_0^T
\nonumber\\
&+&
\int_0^T
\left(
\dot{\boldsymbol{\lambda}}^T
+
\boldsymbol{\lambda}^T\frac{\partial \mathbf{f}}{\partial \mathbf{x}}
+
\frac{\partial h}{\partial \mathbf{x}}
\right)
\frac{\partial \mathbf{x}}{\partial \mathbf{p}}\,dt .
\label{eq:dLdp_parts}
\end{eqnarray}
Assuming that the initial condition does not depend on the parameters,
$\partial \mathbf{x}/\partial \mathbf{p}(0)=0$, we eliminate the dependence on
$\partial \mathbf{x}/\partial \mathbf{p}$ -- which we cannot calculate-- by choosing the adjoint field $\boldsymbol{\lambda}(t)$ such that
\begin{equation}
\dot{\boldsymbol{\lambda}}^T(t)
+
\boldsymbol{\lambda}^T(t)\frac{\partial \mathbf{f}}{\partial \mathbf{x}}\bigl(\mathbf{x}(t),\mathbf{p}\bigr)
+
\frac{\partial h}{\partial \mathbf{x}}\bigl(\mathbf{x}(t),\mathbf{p}\bigr)
=
0,
\qquad
\boldsymbol{\lambda}(T)=0.
\label{eq:adjoint_row}
\end{equation}
Equivalently, in column-vector form,
\begin{equation}
\dot{\boldsymbol{\lambda}}(t)
+
\left(\frac{\partial \mathbf{f}}{\partial \mathbf{x}}\bigl(\mathbf{x}(t),\mathbf{p}\bigr)\right)^T
\boldsymbol{\lambda}(t)
=
-
\left(\frac{\partial h}{\partial \mathbf{x}}\bigl(\mathbf{x}(t),\mathbf{p}\bigr)\right)^T,
\,
\boldsymbol{\lambda}(T)=0.
\label{eq:adjoint_column}
\end{equation}
With this choice, the gradient of the objective functional reduces to
\begin{equation}
\frac{d\alpha}{d\mathbf{p}}
=
\int_0^T
\left[
\frac{\partial h}{\partial \mathbf{p}}\bigl(\mathbf{x}(t),\mathbf{p}\bigr)
+
\boldsymbol{\lambda}^T(t)\frac{\partial \mathbf{f}}{\partial \mathbf{p}}\bigl(\mathbf{x}(t),\mathbf{p}\bigr)
\right]dt.
\label{eq:gradient_final}
\end{equation}

{\it Conceptual problem.}
Equations \eqref{eq:adjoint_column}--\eqref{eq:gradient_final} provide a formally exact expression for the gradient. However, from a physical perspective, the adjoint equation \eqref{eq:adjoint_column} poses three fundamental difficulties.
First, it requires integration backward in time from $t=T$ to $t=0$. Second, the generator of the adjoint dynamics involves the transpose of the Jacobian matrix, $(\partial \mathbf{f}/\partial \mathbf{x})^T$, which has no direct interpretation in terms of the original forward dynamics. Third, the terminal condition $\boldsymbol{\lambda}(T)=0$ must be imposed explicitly.

\begin{figure*}
    \centering
    \includegraphics[width=\textwidth]{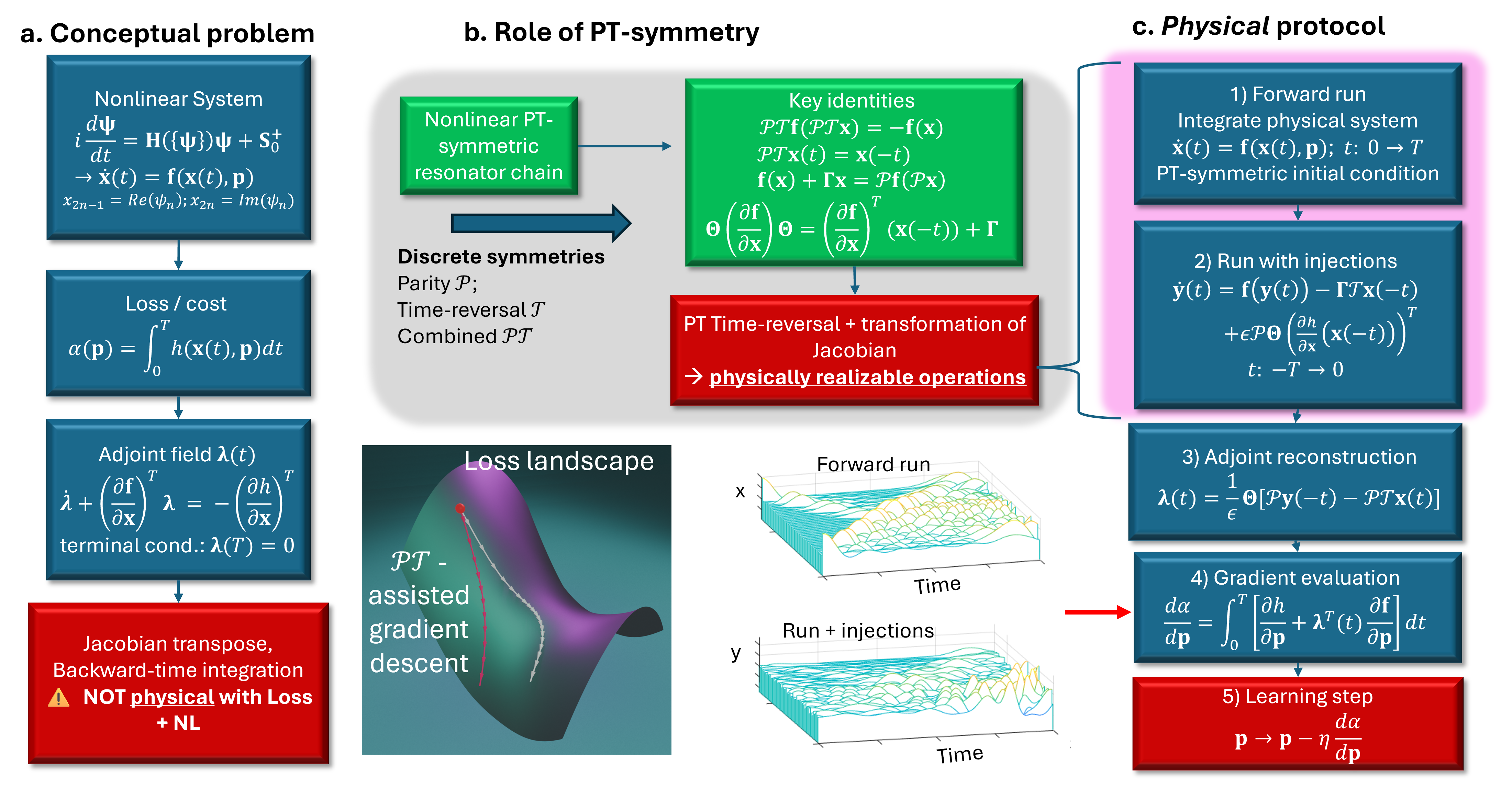} 
    \caption{\textbf{Graphical summary of the $\mathcal{PT}$-assisted physical adjoint optimization protocol.}
    $(a)$ Conceptual obstructions that prevent the direct physical implementation of conventional adjoint optimization in nonlinear, lossy, and nonreciprocal systems.
    $(b)$ In $\mathcal{PT}$-symmetric structures, key identities imposed by the antilinear symmetry map the adjoint dynamics onto physically realizable operations, enabling the adjoint field to be reconstructed from two physical evolutions: a forward run and an injection-driven run.
    $(c)$ Schematic of the resulting physical protocol, consisting of forward evolution, injection-driven evolution, adjoint reconstruction, gradient evaluation, and the subsequent learning step.}
    \label{fig:ExtFig1}
\end{figure*}

In generic nonlinear systems—particularly in non-Hermitian platforms with gain, loss, or non-reciprocal couplings—none of these operations correspond to physically realizable processes. As a result, while the adjoint formalism is mathematically universal, its direct \emph{in-situ} implementation is obstructed at a fundamental level. The remainder of the Methods section is devoted to showing how this obstruction can be removed in nonlinear $\mathcal{PT}$-symmetric systems. See Extended Fig. \ref{fig:ExtFig1} for a graphical summary. 

\subsection{Injection-based realization of tangent dynamics}

Having derived the adjoint equation and identified the fundamental obstruction to its physical implementation, we now consider whether a related, physically accessible dynamics can be constructed as an intermediate step.

A natural starting point is to ask whether the Jacobian itself can be accessed physically. To this end, we introduce an auxiliary trajectory $\mathbf{y}_2(t)$ obeying the same nonlinear dynamics as the original system, but subject to an external injection,
\begin{equation}
\dot{\mathbf{y}}_2(t)=\mathbf{f}(\mathbf{y}_2(t))+\epsilon\,\mathbf{s}(t),
\label{eq:y2_dynamics}
\end{equation}
where $\epsilon\ll 1$ and $\mathbf{s}(t)$ denotes a weak, externally applied driving term,
assumed to be fully controllable and arbitrarily programmable in time within the physical setup.

Assuming that $\mathbf{y}_2(t)$ remains close to $\mathbf{x}(t)$, we expand the vector field to first order,
\begin{equation}
\mathbf{f}(\mathbf{y}_2(t)) \approx \mathbf{f}(\mathbf{x}(t))
+ \left.\frac{\partial \mathbf{f}}{\partial \mathbf{x}}\right|_{\mathbf{x}(t)} \bigl(\mathbf{y}_2(t)-\mathbf{x}(t)\bigr).
\end{equation}
Introducing the deviation $\delta\mathbf{y}(t)=\mathbf{y}_2(t)-\mathbf{x}(t)$, we obtain
\begin{equation}
\delta\dot{\mathbf{y}}(t)
-
\frac{\partial \mathbf{f}}{\partial \mathbf{x}}(\mathbf{x}(t)) \delta\mathbf{y}(t)
=
\epsilon\,\mathbf{s}(t).
\label{eq:tangent_dynamics}
\end{equation}
Equation \eqref{eq:tangent_dynamics} governs the linearized (tangent) dynamics describing the evolution of small deviations around the forward trajectory $\mathbf{x}(t)$, and can be implemented physically through an injection. However, a direct comparison with the adjoint equation~\eqref{eq:adjoint_column} reveals several fundamental discrepancies. First, Eq. \eqref{eq:tangent_dynamics} involves the Jacobian $\partial \mathbf{f}/\partial \mathbf{x}$, rather than its transpose. Second, the sign structure of the evolution differs. Finally, the adjoint equation is subject to a terminal condition $\boldsymbol{\lambda}(T)=0$, whereas Eq.~\eqref{eq:tangent_dynamics} is an initial-value problem.

To partially address these differences, we perform a formal time reversal of the tangent dynamics in Eq. \ref{eq:tangent_dynamics} by defining $\Delta\mathbf{y}(t)=\delta\mathbf{y}(-t)$. This yields
\begin{equation}
\Delta\dot{\mathbf{y}}(t)
+
\frac{\partial \mathbf{f}}{\partial \mathbf{x}}(\mathbf{x}(-t))\,\Delta\mathbf{y}(t)
=
-
\epsilon\,\mathbf{s}(-t).
\label{eq:time_reversed_tangent}
\end{equation}
Equation~\eqref{eq:time_reversed_tangent} is structurally closer to the adjoint equation, and since the driving $\mathbf{s}(t)$ is fully programmable, its functional form can be chosen
freely in an attempt to reproduce the source term appearing in the adjoint equation. We emphasize that this time reversal is, at this stage, a formal manipulation rather than a physically implementable operation.

However, a fundamental obstruction remains: in a generic nonlinear system, the time-reversed Jacobian $\left. \frac{\partial \mathbf{f}}{\partial \mathbf{x}}\right| _{\mathbf{x}(-t)}$ bears no simple relation to the transposed Jacobian $\left(\left. \frac{\partial \mathbf{f}}{\partial \mathbf{x}}\right| _{\mathbf{x}(t)}\right)^T$. No time-independent transformation exists that maps one into the other. This mismatch persists irrespective of the choice of driving $\mathbf{s}(t)$ and cannot be eliminated within conventional nonlinear dynamical systems.

This structural incompatibility motivates the search for systems with additional symmetry constraints. In the following, we show that nonlinear $\mathcal{PT}$-symmetric systems possess precisely the identities required to relate time reversal and Jacobian transposition, thereby enabling a fully physical realization of adjoint dynamics.

\subsection{Nonlinear $\mathcal{PT}$-symmetric resonator chain}
\label{sec:PT_model}

We now introduce a nonlinear $\mathcal{PT}$-symmetric resonator chain that provides the structural identities required to overcome the obstructions identified in Sec. \ref{sec:adjoint_obstruction}. The role of this section is twofold: first, to define a physically realistic nonlinear system in which $\mathcal{PT}$ symmetry is preserved; and second, to establish a set of symmetry identities that allow backward-time evolution and Jacobian transposition to be mapped onto physically accessible operations.

\textit{  Model definition.-}
We consider a one-dimensional chain of $2N$ coupled nonlinear resonators with nearest-neighbor coupling. The dynamics of the complex field amplitudes 
$\boldsymbol{\psi} = (\psi_1,\ldots,\psi_{2N})^T \in \mathbb{C}^{2N}$ is governed by the temporal coupled mode theory equation
\begin{equation}
i\dot{\boldsymbol{\psi}}(t) = \boldsymbol{\mathcal{H}}(\boldsymbol{\psi})\,\boldsymbol{\psi}(t).
\label{eq:psi_dynamics}
\end{equation}
The nonlinear Hamiltonian $\boldsymbol{\mathcal{H}}(\boldsymbol{\psi})$ is tridiagonal and has the form
\begin{equation}
\bigl(\boldsymbol{\mathcal{H}}(\boldsymbol{\psi})\bigr)_{jk}
=
\begin{cases}
i\,\gamma_j + g_j\!\left(|\psi_j|^2\right), & j = k, \\[6pt]
\kappa_j, & k = j+1, \\[4pt]
\kappa_{j-1}, & k = j-1, \\[4pt]
0, & \text{otherwise},
\end{cases}
\label{eq:hamiltonian_components}
\end{equation}
where $\gamma_j$ denotes the linear gain ($\gamma_j>0$) or loss ($\gamma_j<0$) coefficient at site $j$, $g_j(|\psi_j|^2)\in\mathbb{R}$ represents a local nonlinear frequency shift (possibly including a linear frequency offset), and $\kappa_j$ are nearest-neighbor coupling constants.
The system parameters are chosen to satisfy the symmetry constraints
\begin{align}
\gamma_j &= -\gamma_{2N-j+1}, \nonumber\\
g_j(\cdot) &= g_{2N-j+1}(\cdot), \qquad j=1,\ldots,N, \label{eq:PT_conditions}\\
\kappa_j &= \kappa_{2N-j}, \qquad j=1,\ldots,2N-1, \nonumber
\end{align}
to ensure $\mathcal{PT}$ symmetry of the full nonlinear dynamics.

To formulate the adjoint problem, it is convenient to work in real position-momentum coordinates $q_j$ and $p_j$ such that $\psi_j=q_j+ip_j$. We define the real state vector $\mathbf{x}=(q_1,p_1,\ldots,q_{2N},p_{2N})^T\in\mathbb{R}^{4N}$,
for which Eq.~\eqref{eq:psi_dynamics} can be written as
\begin{equation}
\dot{\mathbf{x}}(t)=\mathbf{f}(\mathbf{x}(t)).
\label{eq:real_dynamics}
\end{equation}

\textit{ $\mathcal{PT}$ symmetry in real position-momentum variables.-} 
We now define the parity ($\boldsymbol{\mathcal{P}}$) and time-reversal ($\boldsymbol{\mathcal{T}}$) operators acting on the real position-momentum variables. The parity operator exchanges resonators symmetrically about the center of the chain,
\begin{equation*}
(q_1,p_1,\ldots,q_{2n},p_{2n})
\;\xleftrightarrow{\;\boldsymbol{\mathcal{P}}\;}\;
(q_{2n},p_{2n},\ldots,q_1,p_1),
\end{equation*}
while the time-reversal operator reverses the sign of all momenta,
\begin{equation*}
\boldsymbol{\mathcal{T}}:(q_j,p_j)\mapsto(q_j,-p_j).
\end{equation*}
Their combined action defines the $\boldsymbol{\mathcal{PT}}$ transformation on $\mathbb{R}^{4N}$. 

The nonlinear vector field $\mathbf{f}(\mathbf{x})$ satisfies the $\mathcal{PT}$ symmetry condition
\begin{equation}
\boldsymbol{\mathcal{PT}}\,\mathbf{f}(\boldsymbol{\mathcal{PT}}\,\mathbf{x})=-\mathbf{f}(\mathbf{x}),
\label{eq:PT_vector_field}
\end{equation}
which follows from the parameter constraints Eq. \eqref{eq:PT_conditions}. This identity implies that under the $\boldsymbol{\mathcal{PT}}$ transformation, the system dynamics remains invariant. As a consequence, if the initial condition is $\mathcal{PT}$ symmetric, $\mathbf{x}(0)=\boldsymbol{\mathcal{PT}}\,\mathbf{x}(0)$, then the trajectory satisfies
\begin{equation}
\boldsymbol{\mathcal{PT}}\,\mathbf{x}(t)=\mathbf{x}(-t).
\label{eq:PT_trajectory}
\end{equation}
This property provides a physically meaningful mapping between forward and backward evolution without requiring explicit time reversal.

\textit{  Injection-induced parity symmetry.-} 
For the present system, the vector field also satisfies the identity
\begin{equation}
\boldsymbol{\mathcal{P}}\,\mathbf{f}(\boldsymbol{\mathcal{P}}\,\mathbf{x}) = \mathbf{f}(\mathbf{x})+\boldsymbol{\Gamma}\,\mathbf{x}, 
\label{eq:Gamma_identity}
\end{equation}
where the gain/loss distribution of the chain is encoded in the diagonal matrix $\boldsymbol{\Gamma}=-2\,\mathrm{diag}(\gamma_1,\gamma_1,\ldots,\gamma_{2N},\gamma_{2N})$.
This relation shows that parity-reversed dynamics can be implemented through a controlled linear injection proportional to $\boldsymbol{\Gamma}\,\mathbf{x}$. Importantly, this transformation does not require any modification of the intrinsic nonlinear couplings or gain/loss coefficients, and can be realized physically through external driving.

\textit{  Jacobian transposition via $\boldsymbol{\Theta}$ transformation.-}
The final—and crucial—ingredient concerns the Jacobian of the nonlinear flow,
$\frac{\partial \mathbf{f}}{\partial \mathbf{x}}(\mathbf{x})$. Owing to the nearest-neighbor structure of the chain, $\mathbf{J}(\mathbf{x})$ has a block tridiagonal form. 

We define the matrix
\begin{equation}
\bigl(\boldsymbol{\Theta}\bigr)_{ij}=\delta_{i,4N-j+1},
\label{eq:Theta_def}
\end{equation}
which reverses the ordering of phase-space coordinates by mapping the
position (momentum) coordinate of resonator $m$ onto the momentum
(position) coordinate of its parity-mirror partner $(2N-m+1)$.
Note that $\boldsymbol{\Theta}^2=\mathbf{I}$.
For trajectories satisfying a $\mathcal{PT}$-symmetric initial condition, one can show that the Jacobian obeys the identity
\begin{equation}
\boldsymbol{\Theta}\,\frac{\partial \mathbf{f}}{\partial \mathbf{x}}[\mathbf{x}(t)]\,\boldsymbol{\Theta}
=
\left( \frac{\partial \mathbf{f}}{\partial \mathbf{x}}\right)^{T}[\mathbf{x}(-t)]+\boldsymbol{\Gamma}.
\label{eq:Theta_identity}
\end{equation}
Equation \eqref{eq:Theta_identity} establishes a direct relation between
Jacobian transposition and time reversal, up to a known and controllable
injection term. This identity constitutes the key structural result of
the $\mathcal{PT}$-symmetric construction: it allows the generator of the
adjoint dynamics—typically inaccessible in physical systems—to be
emulated through a physically accessible forward-time evolution
supplemented by external injection.

\textit{  Implications for adjoint realization.-}
Taken together, Eqs. \eqref{eq:PT_vector_field}, \eqref{eq:PT_trajectory}, \eqref{eq:Gamma_identity}, and \eqref{eq:Theta_identity} show that all formal operations appearing in the adjoint equation—backward-time evolution, Jacobian transposition, and terminal constraints—can be systematically mapped onto physically accessible transformations in the nonlinear $\mathcal{PT}$-symmetric resonator chain.

\subsection{Nonlinear $\mathcal{PT}$-symmetry assisted adjoint optimization method.} 

We now construct an explicit injection-based protocol that emulates the
adjoint dynamics using only forward-time evolution of a nonlinear
$\mathcal{PT}$-symmetric resonator chain.

Rather than attempting to implement the adjoint equation Eq.~\eqref{eq:adjoint_column} directly—which is physically inaccessible—we construct a physically realizable surrogate of the adjoint dynamics by combining controlled injection with the symmetry constraints of nonlinear $\mathcal{PT}$-symmetric systems.

\textit{Removal of Jacobian transposition via $\boldsymbol{\Theta}$.-}
The primary obstruction in the adjoint equation is the appearance of the
transposed Jacobian. This can be eliminated using the $\boldsymbol{\Theta}$ identity
Eq.~\eqref{eq:Theta_identity}. We define a transformed adjoint field
\begin{equation}
\boldsymbol{\lambda}(t)=\boldsymbol{\Theta}\,\delta\boldsymbol{\varphi}(t).
\label{eq:lambda_theta}
\end{equation}
Substituting Eq. \eqref{eq:lambda_theta} into the adjoint equation
Eq.~\eqref{eq:adjoint_column} and multiplying from the left by $\boldsymbol{\Theta}$,
we obtain
\begin{equation}
\delta\dot{\boldsymbol{\varphi}}(t)
+
\left[
\frac{\partial \mathbf{f}}{\partial \mathbf{x}}\bigl(\mathbf{x}(-t)\bigr)
-
\boldsymbol{\Theta}\boldsymbol{\Gamma}\boldsymbol{\Theta}
\right]\delta\boldsymbol{\varphi}(t)
=
-\boldsymbol{\Theta}
\left(
\frac{\partial h}{\partial \mathbf{x}}\bigl(\mathbf{x}(t)\bigr)
\right)^T.
\label{eq:varphi_dynamics}
\end{equation}
Equation~\eqref{eq:varphi_dynamics} is now free of Jacobian transposition,
but still evolves backward in time and contains an additional linear
term proportional to $\boldsymbol{\Gamma}$.

\textit{Time reversal and reconstruction of a physical trajectory.-} 
To address the backward-time structure, we interpret $\delta\boldsymbol{\varphi}(t)$ as the
time-reversed deviation between a nearby trajectory $\boldsymbol{\varphi}(t)$ and the
forward solution $\mathbf{x}(t)$,
\begin{equation}
\delta\boldsymbol{\varphi}(t)=\frac{\boldsymbol{\varphi}(-t)-\mathbf{x}(-t)}{\epsilon},
\end{equation}
where $\epsilon\ll1$ ensures that $\boldsymbol{\varphi}(t)$ remains a small deviation from the trajectory $\mathbf{x}(t)$.

Substituting this expression into Eq.~\eqref{eq:varphi_dynamics} and
reconstructing the linearized tangent dynamics yields
\begin{equation}
\dot{\boldsymbol{\varphi}}(t)
=
\mathbf{f}\bigl(\boldsymbol{\varphi}(t)\bigr)
+
\boldsymbol{\Gamma}\,\boldsymbol{\varphi}(t)
-
\boldsymbol{\Gamma}\,\mathbf{x}(t)
+\epsilon
\boldsymbol{\Theta}
\left(
\frac{\partial h}{\partial \mathbf{x}}\bigl(\mathbf{x}(-t)\bigr)
\right)^T,
\end{equation}
where we used $\boldsymbol{\Theta}\boldsymbol{\Gamma}\boldsymbol{\Theta}=-\boldsymbol{\Gamma}$ and restored the nonlinear
vector field $\mathbf{f}(\boldsymbol{\varphi})$.

\textit{Elimination of the gain--loss term via parity injection.-}
The remaining nonphysical element is the gain--loss contribution
$\boldsymbol{\Gamma}\,\boldsymbol{\varphi}(t)$. This term can be removed using the parity-injection
identity Eq.~\eqref{eq:Gamma_identity}. Defining
\begin{equation}
\boldsymbol{\phi}(t)=\boldsymbol{\mathcal{P}}\,\boldsymbol{\varphi}(t),
\end{equation}
we obtain a fully physical forward-time evolution equation
\begin{equation}
\dot{\boldsymbol{\phi}}(t)
=
\mathbf{f}\bigl(\boldsymbol{\phi}(t)\bigr)
+
\boldsymbol{\Gamma}\,\boldsymbol{\mathcal{T}}\,\mathbf{x}(-t)
+
\epsilon\,\boldsymbol{\mathcal{P}}\,\boldsymbol{\Theta}
\left(
\frac{\partial h}{\partial \mathbf{x}}\bigl(\mathbf{x}(-t)\bigr)
\right)^T.
\label{eq:phi_final}
\end{equation}

Equation~\eqref{eq:phi_final} corresponds to the original nonlinear
$\mathcal{PT}$-symmetric dynamics supplemented by a known, controllable
injection term that depends only on the previously measured forward
trajectory $\mathbf{x}(t)$, from $t=0$ to $t=T$.

\textit{Boundary condition and adjoint reconstruction.-} 
The adjoint terminal condition $\boldsymbol{\lambda}(T)=0$ translates into the initial
condition
\begin{equation}
\boldsymbol{\phi}(-T)=\boldsymbol{\mathcal{T}}\,\mathbf{x}(T),
\end{equation}
which can be imposed experimentally.

Once $\boldsymbol{\phi}(t)$ is obtained, the adjoint field is reconstructed as
\begin{equation}
\boldsymbol{\lambda}(t)
=
\frac{\boldsymbol{\Theta}\bigl(\boldsymbol{\mathcal{P}}\,\boldsymbol{\phi}(-t)-\boldsymbol{\mathcal{PT}}\,\mathbf{x}(t)\bigr)}{\epsilon},
\label{eq:adj_reconstruction}
\end{equation}
and the gradient follows directly from Eq. \eqref{eq:gradient_final}.

\subsection{Summary of the protocol}

The complete adjoint-evaluation procedure consists of the following
steps (see Extended Figure \ref{fig:ExtFig1} for a graphical description):

\begin{enumerate}
\item Forward evolution: integrate $\dot{\mathbf{x}}(t)=\mathbf{f}(\mathbf{x}(t),\mathbf{p})$ from $t=0$ to
$t=T$ starting from a $\mathcal{PT}$-symmetric initial condition.
\item Injection-driven evolution: integrate Eq.~\eqref{eq:phi_final} from
$t=-T$ to $t=0$ with initial condition
$\boldsymbol{\phi}(-T)=\boldsymbol{\mathcal{T}}\,\mathbf{x}(T)$.
\item Adjoint reconstruction: compute $\boldsymbol{\lambda}(t)$ using
Eq.~\eqref{eq:adj_reconstruction}.
\item Gradient evaluation: compute $d\alpha/d\mathbf{p}$ using
Eq.~\eqref{eq:gradient_final}.
\end{enumerate}

This protocol realizes the adjoint dynamics entirely through
forward-time evolution of a nonlinear $\mathcal{PT}$-symmetric system,
thereby removing the fundamental physical obstruction identified in
Sec.~\ref{sec:adjoint_obstruction}.

\subsection{Numerical Experiment Protocol}
In our examples, we first solve the forward problem using the DOP853 solver, an eighth-order Runge--Kutta method. We use a relative tolerance of \(10^{-12}\) and an absolute tolerance of \(10^{-15}\), and record the resulting trajectory. We then run the injected simulation using the same solver and tolerances, with injection parameter \(\epsilon = 10^{-5}\).

The optimization variable is the stacked half-chain parameter vector
$\mathbf{p} = \begin{bmatrix} 
\boldsymbol{\gamma}^T  , \boldsymbol{\chi}^T , \boldsymbol{\kappa}^T \end{bmatrix}^T$,
where each of $\boldsymbol{\gamma}$, $\boldsymbol{\chi}$, and $\boldsymbol{\kappa}\in \mathbb{R}^8$, with box constraints
$\chi_j \in [0.001,0.1]$ and $\kappa_j \in [1.0,3.0]$ for all $j=1,\dots,8$.
In each example, seven of the eight gain/loss
coefficients are fixed to zero, while the remaining coefficient is
optimized. Specifically, we set $\gamma_j=0$ for the neutral resonators, and use $\gamma_5 \in [0.05, 0.2]$ for the case where we set the gain/loss indices to be $5$ and $12$, and we use $\gamma_8 \in [0.05, 0.8]$ when gain/loss indices are set to $8$ and $9$. 
Together with the eight independent nonlinear coefficients and eight coupling constants, this yields a
total of 17 independent optimization parameters, whose allowed ranges define the feasible set $\mathcal{B}$.

Since the objective functions considered here are generally highly
nonconvex, we employ a projected gradient method with multiple random
restarts to reduce sensitivity to the initial parameter configuration
and improve exploration of the optimization landscape. For each restart
$r=1,2,\ldots,500$, an initial parameter vector
$\mathbf{p}_0^{(r)}\in\mathcal{B}$
is drawn uniformly at random from the feasible set.


At iteration $k$, the objective function and its adjoint gradient are
evaluated as
\begin{equation}
\alpha_k^{(r)}
=
\alpha\!\left(\mathbf{p}_k^{(r)}\right),
\qquad
\mathbf{g}_k^{(r)}
=
\boldsymbol{\nabla}_{\mathbf{p}}\alpha\!\left(\mathbf{p}_k^{(r)}\right),
\end{equation}
where we define the column gradient as
$\boldsymbol{\nabla}_{\mathbf{p}}\alpha\equiv(d\alpha/d\mathbf{p})^T$.

The trial step size is chosen adaptively according to
\begin{equation}
\eta_k^{\mathrm{raw}}
=
\min\!\left(
\beta\,\eta_{k-1}^{\mathrm{acc}},\,10^{-1}
\right),
\end{equation}
where $\eta_{k-1}^{\mathrm{acc}}$ is the step size accepted at the
previous iteration, $\beta=1.5$, and the initial step size is
$\eta_0^{\mathrm{raw}}=10^{-3}$. A projected trial point is then
constructed as 
$\mathbf{p}_{\mathrm{trial}}^{(r)}
=
\boldsymbol{\Pi}_{\mathcal B}
\left[
\mathbf{p}_k^{(r)} - \eta_k^{\mathrm{raw}}\mathbf{g}_k^{(r)} \right]$,
where $\boldsymbol{\Pi}_{\mathcal B}$ denotes projection onto the feasible parameter
set. Defining the corresponding displacement as $\mathbf{d}_k^{(r)} = \mathbf{p}_{\mathrm{trial}}^{(r)}-\mathbf{p}_k^{(r)}$,
the trial step is accepted if it satisfies the Armijo condition
\begin{equation}
\alpha\!\left(\mathbf{p}_{\mathrm{trial}}^{(r)}\right)
\leq
\alpha_k^{(r)}
+
c_1
\left(\mathbf{g}_k^{(r)}\right)^T \mathbf{d}_k^{(r)}.
\label{eq:armijo}
\end{equation}
If this condition is not satisfied, the step size is successively
reduced by a factor of two and the projected trial point is recomputed,
for up to 25 backtracking steps. Upon acceptance,
$\mathbf{p}_{k+1}^{(r)}=\mathbf{p}_{\mathrm{trial}}^{(r)}$ and the accepted step size is
stored as $\eta_k^{\mathrm{acc}}$. If no acceptable step is found, the
restart is terminated.

Each restart is run for at most 1000 iterations. After completing all
500 restarts, we select the solution with the smallest final objective,
\begin{equation}
r^\star
=
\operatorname*{arg\,min}_{1\leq r\leq500}
\alpha\!\left(\mathbf{p}_{\mathrm{final}}^{(r)}\right),
\qquad
\mathbf{p}^\star=\mathbf{p}_{\mathrm{final}}^{(r^\star)},
\end{equation}
which defines the optimized parameter configuration reported in the
main text.

\end{document}